\documentclass[conference]{IEEEtran}
\IEEEoverridecommandlockouts
\usepackage{cite}
\usepackage{amsmath,amssymb,amsfonts}
\usepackage{algorithmic}
\usepackage{graphicx}
\usepackage{textcomp}
\usepackage{xcolor}
\def\BibTeX{{\rm B\kern-.05em{\sc i\kern-.025em b}\kern-.08em
    T\kern-.1667em\lower.7ex\hbox{E}\kern-.125emX}}

\begin{document}

\title{Reviewing BPMN as a Modeling Notation for CACAO Security Playbooks
\thanks{This research has received funding from the Research Council of Norway (forskningsrådet) under Grant Agreement No. 303585 (CyberHunt project), the EU Connecting Europe Facility (CEF) programme under Grant Agreement No. INEA/CEF/ICT/A2020/2373266 (JCOP project) and the Horizon Europe programme under Grant Agreement No. 101070586 (PHOENi$^2$X project). The views and opinions expressed herein are those of the authors only and do not necessarily reflect those of the European Union or the European Research Council Executive Agency. Neither the European Union nor the granting authority can be held responsible. In addition, this work was supported by the Norwegian R\&I provider CYENTIFIC AS (Organization No. 930271691).
}
}

\author{
\IEEEauthorblockN{Mateusz Zych}
\IEEEauthorblockA{
\textit{University of Oslo}\\
Oslo, Norway \\
mateusdz@ifi.uio.no}
\and
\IEEEauthorblockN{Vasileios Mavroeidis}
\IEEEauthorblockA{
\textit{University of Oslo}\\
Oslo, Norway \\
vasileim@ifi.uio.no}
\and
\IEEEauthorblockN{Konstantinos Fysarakis}
\IEEEauthorblockA{
\textit{Sphynx Analytics Limited}\\
Nicosia, Cyprus \\
fysarakis@sphynx.ch}
\and
\IEEEauthorblockN{Manos Athanatos}
\IEEEauthorblockA{
\textit{Technical University of Crete}\\
Chania, Greece \\
mathanatos@tuc.gr}
}
\maketitle

\begin{abstract}
As cyber systems become increasingly complex and cybersecurity threats become more prominent, defenders must prepare, coordinate, automate, document, and share their response methodologies to the extent possible. The CACAO standard was developed to satisfy the above requirements, providing a common machine-readable framework and schema for documenting cybersecurity operations processes, including defensive tradecraft and tactics, techniques, and procedures. Although this approach is compelling, a remaining limitation is that CACAO provides no native modeling notation for graphically representing playbooks, which is crucial for simplifying their creation, modification, and understanding. In contrast, the industry is familiar with BPMN, a standards-based modeling notation for business processes that has also found its place in representing cybersecurity processes. This research examines BPMN and CACAO and explores the feasibility of using the BPMN modeling notation to represent CACAO security playbooks graphically. The results indicate that mapping CACAO and BPMN is attainable at an abstract level; however, conversion from one encoding to another introduces a degree of complexity due to the multiple ways CACAO constructs can be represented in BPMN and the extensions required in BPMN to support CACAO fully.

\end{abstract}


\section{Introduction}
As the ENISA Threat Landscape 2022 report highlights, cybersecurity attacks continuously grow in number, sophistication, and impact \cite{ENISA-ETL-2022}. The digital transformation, which has accelerated during the COVID-19 pandemic, provides greater opportunities to organizations, governments, and society. However, it results in a rapidly growing complex digital infrastructure with blurred boundaries of responsibilities, multiple suppliers, and challenges to manage. As a result, the negative impact cybersecurity incidents can have on economies, societies, and democracies also grows. 

To enhance cybersecurity across the European Union (EU), the European Commission (EC) has legislated the Network and Information Security (NIS) Directive \cite{NIS} that was adopted in 2016, also focusing on cooperation and collaboration between EU countries by encouraging and, in certain cases, mandating information exchange to jointly tackle emerging security threats and risks. Recently, NIS2 was published, significantly expanding the range of sectors required to achieve a high common level of cybersecurity \cite{NIS-2}. The maturity of defending entities across the globe differs significantly; thus, exchanging knowledge and learning from each other is essential to enhance our cybersecurity capabilities. In fact, recent EU initiatives, such as NIS2, mandate developing cooperation and collaboration mechanisms \cite{NIS-2} to effectively and efficiently respond to cyber incidents, including compliance with reporting obligations. To this end, the Security Operations Centers (SOCs) of Member States and their operators of essential services should incorporate incident response enablers that align with good practices established by the EU authorities and provide the methods and processes needed to guarantee interaction with other SOC instances and associated EU actors. This requires putting in place and following standard operating procedures and having the necessary technology to support their execution and ensure compliance with policies and regulatory frameworks \cite{fysarakis2022blueprint}. However, incident response is a complex function that requires significant planning, resources, and coordination among multiple stakeholders. Thus, to achieve this from a practical perspective, the established approach of exchanging cyber threat intelligence for \textit{shared} threat situational awareness could also be adopted for sharing cybersecurity operations processes, such as incident response playbooks and workflows, to improve the defense capabilities of collaborating entities. 

A standardization work that aims to provide the means to address the above need is OASIS Collaborative Automated Course of Action Operations (CACAO) \cite{CACAO-Security-Playbooks-v2.0} for cybersecurity. CACAO is a specification that defines a common playbook schema and taxonomy to create, document, and share cybersecurity operations playbooks. However, it does not specify a modeling notation for visualization, making it practically challenging to work with. On the other hand, Business Process Model and Notation (BPMN) \cite{BPMN-2.0} offers a standardized and rich notation for visualizing business processes. BPMN 2.0 is a widely used standard with a plethora of tools for creating, editing, and automatically executing BPMN workflows. In fact, it is utilized by many cybersecurity professionals for documenting their operating procedures.

Motivated by the above, this paper examines the use of BPMN modeling notation as a candidate to graphically represent CACAO playbooks and provides a high-level construct mapping between CACAO 2.0 \cite{CACAO-Security-Playbooks-v2.0} and BPMN 2.0 \cite{BPMN-2.0}, thus paving the way for CACAO playbook visualization through BPMN-enabled tools. 

The rest of this paper is structured in the following way. Section \ref{sec:background} presents background information on the state of security playbooks and automation and concisely describes CACAO and BPMN. Section \ref{sec-main} presents and analyses research results, meaning a construct mapping between CACAO and BPMN. Thereafter, we highlight a few limitations regarding the mapping and explain the plan forward. Finally, we end the paper with our conclusions.

\section{Background}
\label{sec:background}
\subsection{Cybersecurity Operations Playbooks and Automation}

Defenders must constantly adapt and respond to incidents requiring better tools, specialized processes, applicable skills and knowledge, time, and human power. The challenge lies in achieving effective and efficient interaction between the various elements of a defense environment to ensure a successful and coordinated incident response. Current cyber-defense practices rely heavily on analysts' speed, expertise, and experience. Unfortunately, human-centered practices do not scale sufficiently to match the velocity, volume, and complexity of current and emerging threats. To overcome slow response times and errors in judgment and ensure a robust, coordinated, and uniform response across multiple teams, defenders standardize and, to the extent possible, automate incident response processes and procedures. A particular focus is given to repetitive tasks, which, if automated, may speed up the response to an incident and allow analysts to perform other more intellectually demanding duties \cite{ENISA-ETL-2022}. In this context, a playbook is devised to document detection, investigation, prevention, mitigation, and remediation steps, capturing managed and repeatable cybersecurity and network operations in support of preparedness, automation, sharing, and interoperability. The execution of a playbook may be fully automated, semi-automated, or manual. Further, playbooks are a fundamental component of Security Orchestration, Automation and Response (SOAR) platforms, which aim to orchestrate software and human agents and automate defensive procedures to the extent possible and according to the guidelines of the underlying playbook executed.

As playbooks can play a significant role in cyber defense, we need a faster and easier way to create, manipulate, and visualize them. In this research, we focus on CACAO, the first open standard for cybersecurity playbooks, which, as for now, does not specify any modeling notation to support graphical representation.

\subsection{Collaborative Automated Course of Action Operations}

\begin{figure}[!ht]
    \centering
    \includegraphics[width=0.3\textwidth]{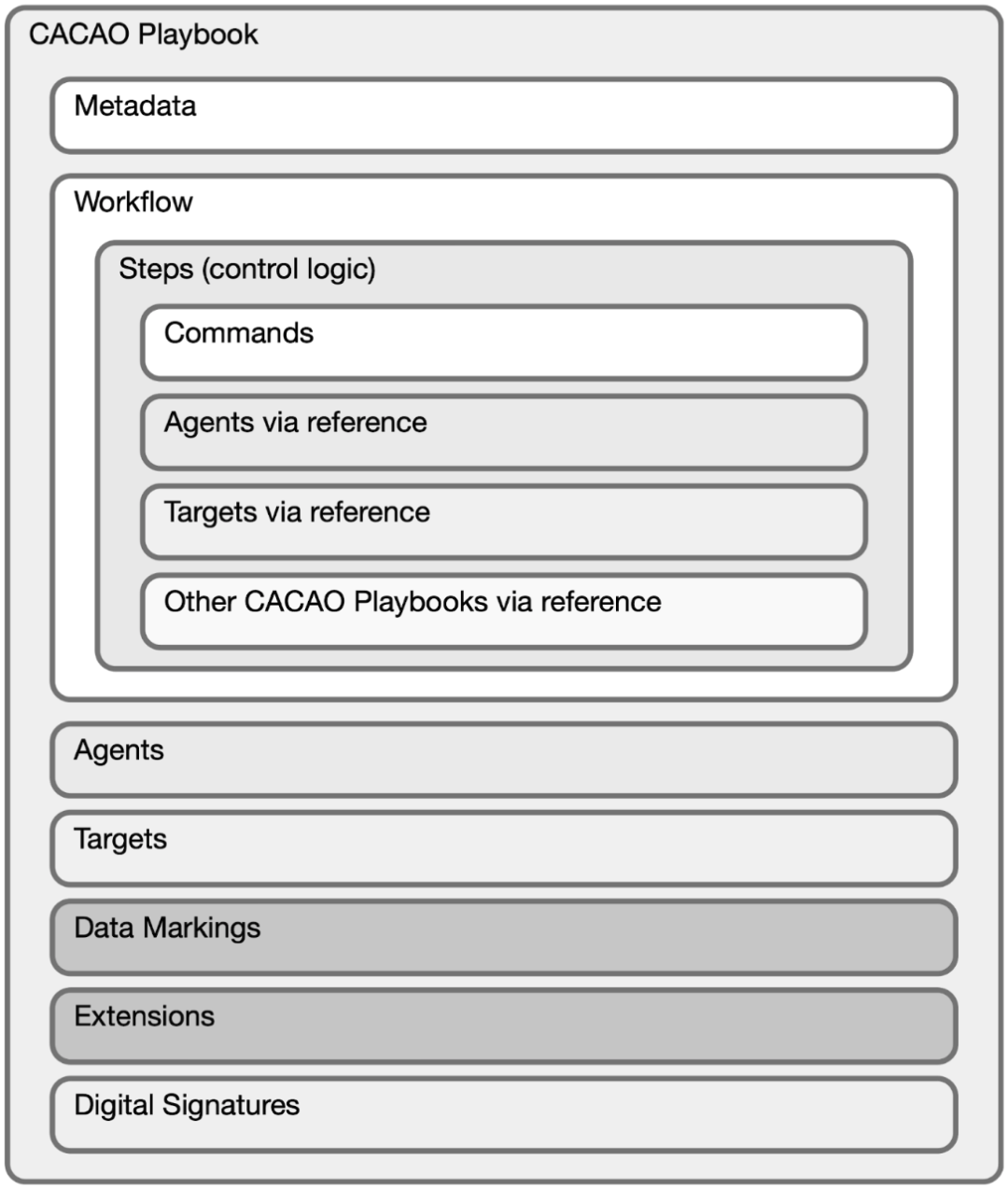}
    \caption{Structure of CACAO 2.0 security playbook \cite{CACAO-Security-Playbooks-v2.0}.}
    \label{fig:cacao_playbook}
\end{figure}

The CACAO specification \cite{CACAO-Security-Playbooks-v2.0} defines a standard schema and taxonomy for creating structured machine-readable cybersecurity playbooks that can be exchanged across organizational boundaries and technical solutions. In support of interoperability reasons and adoption by the global community, CACAO is designed to be vendor-agnostic. Fig. \ref{fig:cacao_playbook} presents the building blocks of a CACAO playbook.

CACAO is created to be flexible in terms of use and built-in automation capability. Thus, adopters, based on their objectives, use cases, and maturity, can create playbooks in different levels of abstraction, detail, and complexity, including the coordination of different operational roles, defense functions, and entities involved in their execution. CACAO is maintained by the OASIS Collaborative Automated Course of Action Operations Technical Committee (CACAO TC) \cite{cacao-tc}. Proof of concept implementations utilizing CACAO have manifested. For example, \cite{mavroeidis2021integration} and \cite{mavroeidis2022cybersecurity} introduced extension mechanisms to relate and ship CACAO playbooks with cyber threat intelligence. \cite{fysarakis2022blueprint} brought up CACAO in a conceptual blueprint for architecting and establishing interoperable SOCs in the EU.

\subsection{Business Process Model and Notation} 
BPMN 2.0 \cite{BPMN-2.0} is a standard for graphically representing business processes in different levels of abstraction, from high-level overviews to detailed executable workflows. With BPMN, organizational processes become more accessible and understandable by both technical and non-technical personnel and gives organizations the ability to communicate these processes in a standardized manner. It is used in various industries, including the field of cybersecurity. Briefly, BPMN provides a rich set of graphical elements representing activities, events, gateways, data objects, messages, and flows and is supported by a wide range of modeling tools and platforms, making it accessible and practical for organizations of all sizes. The BPMN 2.0 standard is maintained by the Object Management Group (OMG), a standards development organization. The last version of the standard, BPMN 2.0, was published in 2011, and in 2013, BPMN 2.0 became an international standard through ISO/IEC 19510 \cite{ISO19510}.

\section{A BPMN-based Modeling Notation for CACAO Playbooks}
\label{sec-main}

CACAO playbooks are serialized in JavaScript Object Notation (JSON) format. Although JSON is both human- and machine-readable, even a non-complex CACAO playbook will consist of hundreds of lines of code, making it challenging for a defender to design, maintain, comprehend, and utilize it. As CACAO is still in its early adoption phase, no modeling notation has yet been considered by the CACAO TC. Providing the ability to graphically represent CACAO playbooks using a standardized notation, like BPMN, will significantly improve defenders' efficiency in generating, maintaining, and executing playbooks. 
This research aims to break this barrier and provides a mapping between CACAO 2.0 constructs and BPMN 2.0 shapes/constructs to incorporate a modeling notation to CACAO and allow for a two-way conversion.

Producers and consumers of CACAO 2.0 must conform to the requirements defined in section 10 of the specification \cite{CACAO-Security-Playbooks-v2.0}. Considering the mandatory and optional CACAO features to support, this work identifies BPMN constructs with the same meaning, aiming to use a subset of the modeling notation to graphically represent CACAO playbooks. Per the CACAO 2.0 specification, producers and consumers must support versioning, variables, playbooks, workflow steps, commands, agents, and targets. Optional features are data markings, extensions, and digital signatures.

\begin{figure*}
   \centering
       \includegraphics[width=1\textwidth]{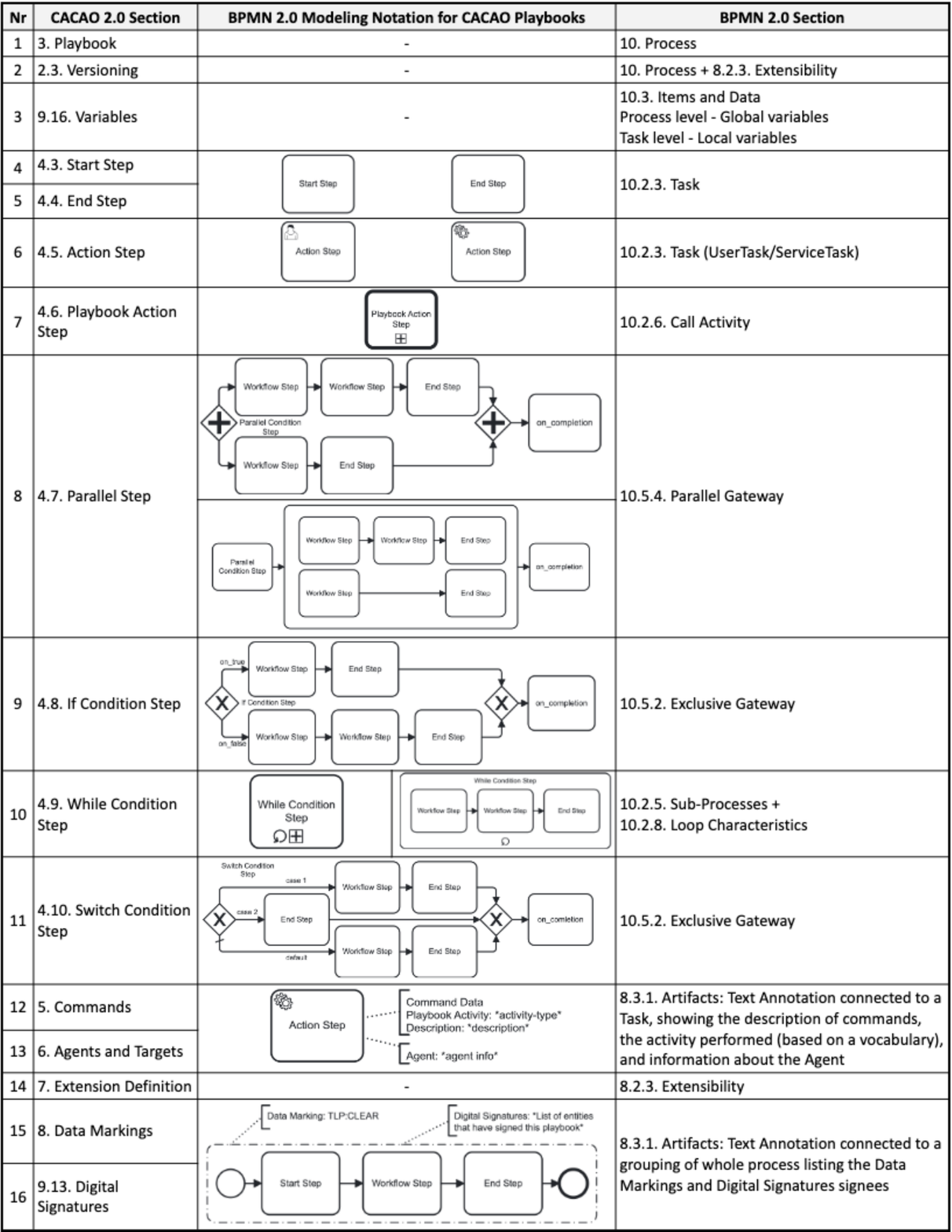}
 \caption{Construct mapping between CACAO 2.0 \cite{CACAO-Security-Playbooks-v2.0} and BPMN 2.0 \cite{BPMN-2.0}.}
 \label{fig:cacao-bpmn-mapping}
\end{figure*}

The subsequent sections cover the CACAO features producers and consumers must support and their counterpart constructs in BPMN, followed by the optional features. Finally, all construct mappings between CACAO 2.0 and BPMN 2.0 are presented in Fig. \ref{fig:cacao-bpmn-mapping}\footnote{https://github.com/cyentific-rni/bpmn-cacao}.

\subsection{Playbook}
In BPMN, a Process is equivalent to a Playbook in CACAO (Fig. \ref{fig:cacao-bpmn-mapping}, row 1). A BPMN Process (i.e., playbook/workflow) comprises a set of Activities, Events, Gateways, and Sequence Flows. A BPMN Process can be reused by other Processes, as CACAO playbooks can be reused by other CACAO playbooks (see subsection \ref{playbook-action-step}). The Playbook object in CACAO, as presented in Fig. \ref{fig:cacao_playbook}, is a wrapper for all CACAO classes of objects/constructs encompassing a CACAO Playbook.

\subsection{Metadata}
An important feature of CACAO playbooks is their rich cybersecurity-focused metadata (CACAO 2.0, section 3 \cite{CACAO-Security-Playbooks-v2.0}) that can be used to inform how the execution of a playbook should be handled or support knowledge management. Understanding that BPMN was not created with cybersecurity in mind, we have to extend BPMN using the attribute extension mechanism (BPMN 2.0, section 8.2.3 \cite{BPMN-2.0}) on the respective BPMN construct to achieve a complete construct-to-construct conversion. 

However, to fulfill the CACAO conformance requirements, the producer or/and consumer must at least support the metadata pertinent to playbook versioning (Fig. \ref{fig:cacao-bpmn-mapping}, row 2). In particular, versioning combines a playbook identifier with created and modified timestamps to distinguish different versions of the same playbook.

\subsection{Variables}
Two types of Variables exist in CACAO. On the Playbook level, we have global Variables, and on the Workflow Step level, we have local Variables. Correspondingly, BPMN operates with both global and local Variables. For instance, the BPMN globally accessible Variable can be created with an Item Definition element (Fig. \ref{fig:cacao-bpmn-mapping}, row 3) within a Process or a Sub-Process object definition. Then, all elements within the Process can access the globally defined Variable. Similarly, a local Variable for a Task can be created.

\subsection{Workflow Step}
Workflow Steps are the main building blocks of a CACAO playbook. They indicate the Start and End of a playbook, encode logical operations, can call other playbooks, and capture a sequenced list of commands to be performed by one or several actuators (Agent-Target types). All Workflow Steps in CACAO share a set of properties and may define other unique properties. The following subsections present the mapping of each CACAO Workflow Step type to BPMN 2.0.

\subsubsection{Start and End Step} \label{start-end-step}
in CACAO the Start Step is the starting point of a playbook and represents an explicit entry in the workflow to signify the start of a playbook. The End Step represents an explicit point in the workflow to signify the end of a playbook or branch of steps (see Fig. \ref{fig:cacao-bpmn-mapping}, rows 4 and 5).

In BPMN, the Start and End elements are Events that indicate the Start and the End of a Process. Events can be invoked by triggers, such as a message, timer, or signal. In contrast, as explained above, CACAO 2.0 does not incorporate the concept of events and triggers. Consequently, the semantic difference at the construct level between the two standards makes it infeasible to achieve a mapping, as they play different roles.

In this context, a solution to outmaneuver the aforementioned limitation is to map the CACAO Start and End Steps to BPMN Tasks; one to indicate the Start of the Process and a second to indicate the End of the Process. In addition, according to the BPMN 2.0 specification, the Start and End Events are optional, allowing us to omit them, although it is not a best practice.
  
\subsubsection{Action Step} \label{action-step}

in CACAO, an Action Step is an individual node in a workflow that contains the commands to be performed by an agent against a set of targets (see subsection \ref{agents-and-targets}). Mapping a CACAO Action Step to BPMN is straightforward (see Fig. \ref{fig:cacao-bpmn-mapping}, row 5), as it relates directly to a Task which is an atomic (can not be broken to a finer level of detail) Activity within a Process Flow. In addition, we propose the use of BPMN Text Annotations for presenting the end user details regarding the Commands and the Agent and Target(s) that will be involved in their execution (see rows 12 and 13 in Fig \ref{fig:cacao-bpmn-mapping}). Finally, based on the type of Agent and Target, meaning as per the CACAO specification, people and places or devices and equipment, the Task will be either a User or a Service Task.

\subsubsection{Playbook Action Step} \label{playbook-action-step}
in CACAO a Playbook Action Step references another playbook to be invoked for execution. The associated construct in BPMN is Call Activity, which references a process that is external to the current process definition. This mechanism permits the creation of a global process and/or task definitions that can be reused (see Fig. \ref{fig:cacao-bpmn-mapping}, row 7). 

\subsubsection{Parallel Step} \label{parallel-step}
the Parallel Step in CACAO differs slightly from the identified BPMN Parallel Gateway (see Fig. \ref{fig:cacao-bpmn-mapping}, row 8). They are both used to create concurrency in a Process, but the CACAO Parallel Step is used only to fork and not to join paths of execution. Further, CACAO expects all branching paths of execution to finish by indicating an explicit End Step on each branch, then return to the Parallel Step before continuing to the next Step identified in the construct. These differences in the execution of Parallel Steps between CACAO and BPMN make it challenging to perceive them as a one-to-one mapping since there are several ways to model a set of elements to have a similar execution flow. One interesting option to consider is that of Expanded Embedded Sub-Process used as a “Parallel Box” as shown in Fig. \ref{fig:cacao-bpmn-mapping}, in the bottom of row 8, which conveys similar semantics as the Parallel Step in CACAO.

\subsubsection{If Condition Step} \label{if-condition-step} 
the CACAO If Condition Step represents a regular if-logic as known from various programming languages (see Fig. \ref{fig:cacao-bpmn-mapping}, row 9). However, in CACAO, when the if-statements evaluate to true or false, they create a new branch with at least one Workflow Step and an explicit End Step. Traversing the full path of the branch, the Workflow run will eventually need to move back to the branching point (If Condition Step) to look for the next Step to execute. In BPMN, an Exclusive Gateway represents the if-logic, and when executed, one path is chosen and continues to the end of the Process. Sometimes, the Exclusive Gateway is used to join the if-logic branches; however, only one path is followed during the execution.

We have identified multiple ways to model this type of logic. One way is to include the Exclusive Gateway for the if-branching, including CACAO End Step(s), and leave unutilized the BPMN End Event, then include the Exclusive Gateway for joining the branches before the entity executing the playbook continues to the next Step (see Fig. \ref{fig:cacao-bpmn-mapping}, row 9. Another way is to utilize an Embedded Sub-Process with the if-logic inside and the "on\_true" and "on\_false" branches containing the CACAO End Step and (optionally) the BPMN End Event. Then, when the Embedded Sub-Process completes, the entity executing the playbook will continue with the next Step. 

\subsubsection{While Condition Step} \label{while-condition-step}
the While Condition Step evaluates a condition but only with the option of defining a new branch when it is evaluated to be true. Otherwise, the run continues to the next Step defined in the While Condition Step construct properties. BPMN does not include a distinct modeling notation for a while-loop. Instead, the modeling can be achieved by combining a set of Elements or our preferred method using the Embedded Sub-Process construct with loop characteristics (see Fig. \ref{fig:cacao-bpmn-mapping}, row 10). 

\subsubsection{Switch Condition Step} \label{switch-condition-step}
the Switch Condition Step is a logical construct that can be modeled using the Exclusive Gateway in BPMN with multiple outgoing Sequence Flows. The run will follow only one path. Like the other Steps described in the sections above, we found similar constraints in mapping this Step. According to the CACAO specification, each case in the Switch Condition Step creates a branch of Workflow Steps, including End Step. After the execution of a branch, the entity executing the playbook needs to track the Switch Condition Step back to continue with the next Workflow Step. The dilemma is recurrent. One must avoid using End Events and force a connection between the End Step and the Exclusive Gateway to merge branches before continuing to the next Step (a BPMN-aligned approach) as shown in Fig. \ref{fig:cacao-bpmn-mapping}, row 11. Otherwise, all branches and the diverging Exclusive Gateway must be encapsulated within an Embedded Sup-Process.

\subsection{Commands} \label{commands}
The Commands are an integral part of the CACAO Action Step; thus, no modeling counterpart in BPMN has been identified. However, the Command information is essential as it eases the understanding of a playbook. Therefore, we have decided to utilize the Text Annotation Artifact (BPMN 2.0 section 8.3.1) \cite{BPMN-2.0} to show the description of command(s) and the activity type they perform (e.g., scan for vulnerabilities or configure systems). The Text Annotation has a fixed header, "Command Data," as shown in Fig. \ref{fig:cacao-bpmn-mapping}, row 12.

\subsection{Agents and Targets} \label{agents-and-targets}
The CACAO Agent-Target data type represents information about entities or devices that accept, receive, process, or execute one or more commands defined in an Action Step \cite{CACAO-Security-Playbooks-v2.0}. This data type is used by two different properties at the Playbook level of CACAO, namely, agent\_definitions and target\_definitions, and are utilized/referenced in Action Steps. In addition, as explained in section \ref{action-step}, the Agent-Target data type determines the BPMN Task type to which an Action Step is modeled/mapped. For example, an Action Step will utilize a User Task for Agents representing people and places, whereas a Service Task applies only to devices and equipment. Although this data type does not have a distinct modeling element, it is essential to graphically represent that information to understand the mechanics of a Playbook. Therefore, we show this information to the user using the BPMN Text Annotation Artifact in connection to the CACAO Action Step/BPMN Task (see Fig. \ref{fig:cacao-bpmn-mapping}, row 13).

\subsection{Extension Definition} \label{extension-definition}
Both BPMN and CACAO allow extending the base specifications by enriching existing constructs or defining new ones. In BPMN, enriched/extended constructs utilize the same modeling elements (shapes). In contrast, new constructs should define their representation and must not conflict with existing elements in BPMN (see Fig. \ref{fig:cacao-bpmn-mapping}, row 14).

\subsection{Data Markings} \label{data-markings}
Data markings provide essential information about the handling and/or sharing requirements of a playbook. They are primarily informative but can be acted upon but are not part of the Process Flow; therefore, no dedicated BPMN element has been identified. However, we represent it visually by placing the whole Process in a Grouping with a Text Annotation Artifact displaying the Data Markings used (see Fig. \ref{fig:cacao-bpmn-mapping}, row 15).

\subsection{Digital Signatures} \label{digital-signatures}
In CACAO, Digital Signatures, like Data Markings, do not alter the playbook, nor do they contribute to the work performed by the Workflow Steps. However, this information is valuable and can be acted upon as it informs the consumer about the entities that have signed the playbook, which is equivalent to indicating levels of trust in it. As in Data Markings, we implement the same approach and display all the parties who have signed the playbook using a Text Annotation Artifact connected to a Grouping encapsulating the whole Process (see Fig. \ref{fig:cacao-bpmn-mapping}, row 16).

\section{Discussion}
\subsection{Limitations}
Several intricacies emerged when creating a construct mapping between BPMN and CACAO. A key consideration is that, in some cases, the mappings are not one-to-one, as some CACAO constructs can be modeled in several ways in BPMN. Another concern is that using Sub-Processes to solve the CACAO construct translation/conversion case mentioned above creates another case, meaning we need several different types of Sub-Processes that will share the common metadata of CACAO Workflow Steps. However, the object-specific (unique) properties will differ. A Parallel Step needs to understand specific properties applicable to the Parallel Conditional Step, whereas the Switch Condition Step needs other specific properties for its logical functions. The above intricacies complicate the creation of an explicit conversion mechanism between these two specifications since the converter would need to look for several different patterns of groups of BPMN elements to map to one construct in CACAO.

\subsection{Future work}
This research is the first step in introducing a modeling notation for CACAO playbooks. In this context, we examined whether we can utilize BPMN. Analyzing the mapping of constructs between the two standards and knowing the intricacies involved enables us to move toward developing software that can convert CACAO playbooks to BPMN workflows and the opposite. While developing the software, a key consideration must be restricting the amount of valid BPMN representations a CACAO construct can be represented, allowing the development of an MVP while efficiently managing development resources. Finally, to achieve a complete conversion, all the utilized BPMN constructs must be extended to accommodate all the metadata of CACAO.

\section{Conclusion}
This paper presented and analyzed a mapping between the CACAO 2.0 and BPMN 2.0 specifications, validating the feasibility of using BPMN to graphically represent CACAO Security Playbooks. Such an approach will be of great value for defenders, as incorporating a modeling notation to CACAO will help to dramatically decrease the time spent comprehending, generating, and maintaining playbooks.

\section*{Acknowledgment}
The authors would like to express their gratitude to Charles Frick from Johns Hopkins University Applied Physics Laboratory and Francisco Luis de Andrés Pérez for their valuable input and discussions.

\bibliographystyle{IEEEtran}
\bibliography{./main}

\end{document}